\documentclass[epjST]{svjour}
\usepackage{graphicx}
\usepackage{amsmath,amssymb}
\begin{document}
\title{Radial vortex core oscillations in Bose-Einstein condensates}
\author{N. Verhelst\inst{1}\fnmsep\thanks{\email{nick.verhelst@uantwerpen.be}} \and T. Ichmoukhamedov\inst{1} \and J. Tempere\inst{1,2}}
\institute{TQC, Universiteit Antwerpen, Universiteitsplein 1, B-2610 Antwerpen, Belgium \and Lyman Laboratory of Physics, Harvard University}
\abstract{
Dilute ultracold quantum gases form an ideal and highly tunable system in which superfluidity can be studied. Recently quantum turbulence in Bose-Einstein condensates was reported [PRL \textbf{103}, 045310 (2009)], opening up a new experimental system that can be used to study quantum turbulence. A novel feature of this system is that vortex cores now have a finite size. This means that the vortices are no longer one dimensional features in the condensate, but that the radial behaviour and excitations might also play an important role in the study of quantum turbulence in Bose-Einstein condensates. In this paper we investigate these radial modes using a simplified variational model for the vortex core. This study results in the frequencies of the radial modes, which can be compared with the frequencies of the thoroughly studied Kelvin modes. From this comparison we find that the lowest (l=0) radial mode has a frequency in the same order of magnitude as the Kelvin modes. However the radial modes still have a larger energy than the Kelvin modes, meaning that the Kelvin modes will still constitute the preferred channel for energy decay in quantum turbulence.
}
\maketitle
\section{Introduction}
\label{intro}

Quantum turbulence\cite{Barenghi3} (QT) was achieved in ultracold atomic gases\cite{Anderson} by shaking the condensate in conjunction with rotation\cite{Henn} or by rapidly sweeping a laser beam through the condensate\cite{Kwon,White}. Since ultracold quantum gases are highly tuneable \cite{Bloch} in comparison to superfluid helium (interaction strength, number of particles, type of trapping, dimensionality, \dots), the realization of QT in these gases opens up new opportunities for this field of study. One example of the new opportunities offered by quantum gases is the fact that it is now possible to investigate QT in a two dimensional (2D) Bose-Einstein condensate (BEC) \cite{Wilson}. 

Before the experimental realization of ultracold gases, QT was extensively studied in superfluid helium \cite{Barenghi,Fisher,Barenghi2}, where it was discovered already 50 years ago \cite{Hall}. In these studies it became apparent that QT is characterized by the appearance of singly quantized vortices that are distributed in a tangled way \cite{Barenghi,Vinen,Kobayashi}. These vortices appear as a consequence of the Kolmogorov decay \cite{Kolmogorov} which causes larger eddies to break up into singly quantized vortices. How the energy further dissipates from the tangled series of quantized vortices has not yet been experimentally observed. It is theorized that the further dissipation of these tangled vortices happens via vortex reconnections and ultimately the excitation of the axial Kelvin waves \cite{Vinen2,Thomson,Thomson2} of the vortex line, leading to an energy dissipation via phonons and rotons\cite{Vinen2,Vinen3,Kondaurova}. These Kelvin waves were also experimentally observed in a lattice of vortices in a BEC \cite{Smith}.

In superfluid helium the size of the vortex core is in the order of nanometers\cite{Donnelly}. In order to get an observation of the vortex flow experimentally micron-sized\cite{Bewley} and even sub-micron sized\cite{Fonda} solid hydrogen trackers are used. In ultracold gases however the vortex core has a size in the order of fractions of micrometers \cite{Cooper}, which yields a typical condensate size to vortex core size ratio of 10-50. Observing a vortex core in ultracold gases is typically done by imaging the condensate after expansion, using tomographical methods \cite{Rosenbusch} for 3D reconstruction, or in situ methods \cite{Gemelke}.

In the present article, the starting idea is the fact that in ultracold atomic gases the size (healing length) of the vortex core is not negligible with respect to the condensate size. The fact that the size of a vortex core in an ultracold gas is non-negligible, yields new effects for the dissipation of energy in the mechanism of QT. The excitations for finite-sized vortex cores that are explored in this article are the radial oscillation modes. This oscillation mode yields an additional way, next to Kelvin modes, for energy to be dissipated by the movement of the vortex core.

\section{Theoretical description}
\label{sec:theory}

In order to describe Bose-Einstein condensates (BECs) the standard Gross-Pitaevskii theory \cite{Gross,Pitaevskii} is used. The condensate wave function $\Psi(\textbf{r},t)$ is described in the hydrodynamical picture:
\begin{equation}
\Psi(\textbf{r},t)=\sqrt{n(\textbf{r},t)}\exp\left(i S(\textbf{r},t)\right),
\label{eq:HydrodynamicalPicture}
\end{equation}
using the real valued density $n(\textbf{r},t)$ and phase $S(\textbf{r},t)$ fields. To consider density fluctuations, this description is most useful. The velocity field is obtained from the phase field using $\textbf{v}(\textbf{r},t)=\frac{\hbar}{m}\nabla S(\textbf{r},t)$. The magnitude of the velocity vector field $\textbf{v}(\textbf{r},t)$ is denoted by the scalar field $v(\textbf{r},t)$.

The system that will be studied is a BEC in a cylindrical box potential with height $H$ and radius $R_\mathrm{cond}$. These homogeneous BECs can also be realized in experiments\cite{Gaunt}. Using the hydrodynamic description, the kinetic and interaction energy of the homogeneous superfluid confined in $r\in[0,R_\mathrm{cond}]$, $z\in[0,H]$ are given by respectively\cite{Pethick}:
\begin{equation}
\begin{aligned}
E_\mathrm{kin}[\Psi]&=\frac{m}{2}\int d^3\textbf{r}\left[n(\textbf{r},t)v^2(\textbf{r},t)\right],\\
E_\mathrm{int}[\Psi]&=\frac{1}{2}g\int d^3\textbf{r}(n_\infty-n(\textbf{r},t))^2,
\end{aligned}
\label{eq:GrossPitaevskiiEnergy}
\end{equation}
where $g=\frac{4\pi\hbar^2a_s}{m}$, with $a_s$ the s-wave scattering length and $n_\infty$ the homogeneous bulk density. In order to calculate the different energies an analytical variational form is used for the superfluid density $n(\textbf{r},t)$ and the velocity field $\textbf{v}(\textbf{r},t)$. Important to note is that the energy equations \eqref{eq:GrossPitaevskiiEnergy} only hold when the condensate size $R_\mathrm{cond}$ is large enough compared to the condensate healing length $\xi$.

\subsection{Describing the single-vortex structure (unperturbed)}
\label{subsec:singlevortexstructure}

Since the vortex core structure $n(\textbf{r})$ has no analytical solution, even in the simplest (unperturbed, cylindrical symmetric) case, a variational function is commonly used in order do derive analytical results. A commonly used variational model is the hyperbolic tangent which yields an accurate fit for the vortex core structure\cite{Verhelst}. In this work a simple cylindrical hole of the form\cite{Ginzburg}
\begin{equation}
n(\textbf{r})=n_\infty\Theta(r-R_\mathrm{v})\Theta(R_\mathrm{cond}-r)\Theta(z)\Theta(H-z)
\label{eq:SingleVortexUnperturbed}
\end{equation}
is used, where $R_\mathrm{v}$ is the size of the vortex core. This model yields the same energy as the hyperbolic tangent variational model up to an additive constant\cite{Fetter}. This means that the variational model \eqref{eq:SingleVortexUnperturbed} yields accurate results if\footnote{A factor of $R_\mathrm{cond}/R_\mathrm{v}$ equal to 10 is already sufficient in the homogeneous case for an error in energy smaller than 1\%.} $R_\mathrm{cond}>>R_\mathrm{v}$. This has only been investigated for the (radial) ground state configuration, and not for the excitations. Nevertheless, the simplicity of the model allows for analytic solutions that can serve as a benchmark for future improvements of the model. Note that also $H>>\xi$ for the variational model \eqref{eq:SingleVortexUnperturbed}, otherwise also the edge effects in the $z$ direction will play a more important role. The velocity field of a single (singly quantized) vortex is given by:
\begin{equation}
\textbf{v}_\mathrm{v}(\textbf{r})=\frac{\hbar}{mr}\textbf{e}_\theta.
\label{eq:VelocityFieldSingleVortex}
\end{equation}

The variational parameter for the above vortex model \eqref{eq:SingleVortexUnperturbed} is the vortex radius $R_\mathrm{v}$. Minimizing the energy of the non-perturbed vortex it can be shown that a minimal energy is achieved when:
\begin{equation}
R_\mathrm{v}=\frac{1}{\sqrt{4\pi n_\infty a_s}}=\sqrt{2}\xi,
\label{eq:MinimalRadiusVortexCore}
\end{equation}
where $\xi$ is the healing length of the condensate. Around this energy minimum all of the considered radial vortex core oscillations will be expanded.

\subsection{Adding perturbations to the single-vortex structure}
\label{subsec:singlevortexstructureperturbed}

Combining the fact that the flow of a BEC is irrotational\cite{Pitaevskii2} ($\nabla\times\textbf{v}(\textbf{r},t)=0$) and that the variational model \eqref{eq:SingleVortexUnperturbed} leads to an incompressible flow ($\nabla\cdot\textbf{v}(\textbf{r},t)=0$), the equation of motion for the BEC velocity field can be derived. The BEC velocity can be described by a velocity potential $\phi(\textbf{r},t)$, obeying a potential equation $\nabla^2\phi(\textbf{r},t)=0$.

Solving the above potential equation in cylindrical coordinates, using the boundary condition $\textbf{v}(r\rightarrow\infty,t)=0$, results in the velocity field is given by:
\begin{equation}
\textbf{v}_l(\textbf{r},t)=\dot{A}_l(t)\left(\frac{r}{R_\mathrm{v}}\right)^{-l-1}\left[\cos(l\theta)\textbf{e}_r+\sin(l\theta)\textbf{e}_\theta\right]\qquad\mathrm{with}\qquad l\in\mathbb{N},
\label{eq:VelocityFieldPerturbation}
\end{equation}
where the integer $l$ labels the different oscillation modes and $\dot{A}_l(t)$ is the integration constant (fixing the size of the $l^\mathrm{th}$ mode). In principle\footnote{For a general velocity field consisting of a combination of different perturbation modes this is needed in order to have a complete Fourier series in the angular coordinate $\theta$.} we should have a linear combination of $\cos(l\theta)$ and $\sin(l\theta)$ for the angular dependency. However for our calculations the relative phase between the different $l$ modes will not play a role since we will be looking at small perturbations. The modes can thus be seen as a set of non interacting\footnote{The cross terms ($\propto\dot{A}_l(t)\dot{A}_k(t)$) in the energy drop out in the calculations.} harmonic oscillators. Important to note is that the true boundary conditions for the velocity field of the system is given by $\textbf{v}(r\rightarrow R_\mathrm{cond},t)=0$, however in practice the value of $R_\mathrm{cond}$ is chosen to be large enough so that the boundary condition mentioned above can be implemented.

\subsection{Describing the radial oscillations}
\label{subsec:radialoscillations}

In order to now introduce the radial oscillations of the vortex core, the vortex core radius (for a given mode $l$) is written as:
\begin{equation}
R_l(\theta,t)=R_\mathrm{v}+\delta R_l(\theta,t),
\label{eq:VortexCoreRadiusExpansion}
\end{equation}
with $R_\mathrm{v}$ the equilibrium case \eqref{eq:MinimalRadiusVortexCore} and $\delta R_l(\theta,t)$ the deformation of the vortex core due to the $l^\mathrm{th}$ oscillation mode. The deformation of the vortex core $\delta R_l(\theta,t)$ can be derived from the velocity field using:
\begin{equation}
\dot{R}(\theta,t)=\delta\dot{R}(\theta,t)=\textbf{v}(R_\mathrm{v},\theta,t)\cdot\textbf{e}_r.
\label{eq:RadiusVelocityFieldVariation}
\end{equation}
The velocity field in \eqref{eq:RadiusVelocityFieldVariation} is evaluated in $r=R_\mathrm{v}$, this can be done since for the perturbations $r=R_\mathrm{v}+\delta R_l$ and $R_\mathrm{v}>>\delta R_l$. 

Since for the stable vortex, the velocity field lies along $\textbf{e}_\theta$, this flow will not deform the vortex core. Using the velocity field \eqref{eq:VelocityFieldPerturbation}, the deformation of the vortex core is given by \eqref{eq:RadiusVelocityFieldVariation}:
\begin{equation}
\delta R_l(\theta,t)=A_l(t)\cos(l\theta).
\label{eq:VortexCoreVariationlthMode}
\end{equation}
Combining \eqref{eq:VortexCoreVariationlthMode} and \eqref{eq:VortexCoreRadiusExpansion} then yields the deformation of the vortex core:
\begin{equation}
R_l(\theta,t)=R_\mathrm{v}+A_l(t)\cos(l\theta).
\label{eq:VortexCoreRadiusPerturbed}
\end{equation}

\section{Results}
\label{sec:results}

Using the description of section \ref{sec:theory} for the perturbed vortex core with radial oscillations, the total energy can be calculated using equations \eqref{eq:GrossPitaevskiiEnergy}. Given the total energy it is possible to derive the frequencies of the different oscillation modes. The obtained frequencies can then be compared with the known frequencies of the well studied Kelvin modes.

\subsection{Energy and frequencies of the radial oscillation modes}
\label{subsec:EnergyAndFrequencies}

The total energy of the vortex core, containing radial perturbations in the vortex core radius can now be calculated by substituting \eqref{eq:SingleVortexUnperturbed} in \eqref{eq:GrossPitaevskiiEnergy}, using \eqref{eq:VortexCoreRadiusPerturbed} for the vortex radius $R_\mathrm{v}$. Since the mode $l=0$ (the breathing mode) will lead to logarithmic divergences, this energy should be calculated separately. Calculating the energies of the vortex core with a radial perturbation yields:
\begin{equation}
\begin{aligned}
E_{l=0}[\Psi]&=E^{(0)}+\frac{\hbar^2\pi n_\infty}{2m\xi^2}HA_0^2(t)+2\xi^2mn_\infty\pi H\ln\left(\frac{R_\mathrm{cond}}{R_\mathrm{v}}\right)\dot{A}^2_0(t),\\
E_{l\neq 0}[\Psi]&=E^{(0)}+\frac{\hbar^2\pi n_\infty}{4m\xi^2}HA_l^2(t)+\frac{1}{l}\xi^2mn_\infty\pi H\left[1-\left(\frac{R_\mathrm{v}}{R_\mathrm{cond}}\right)^{2l}\right]\dot{A}^2_l(t),\\
\end{aligned}
\label{eq:EnergyVortexPerturbed}
\end{equation}
where $E^{(0)}$ is the energy of the unperturbed vortex:
\begin{equation}
E^{(0)}=\frac{\hbar^2\pi n_\infty}{m}H\ln\left(\frac{R_\mathrm{cond}}{R_\mathrm{v}}\right)+\frac{1}{2}\frac{4\pi\hbar^2a_s}{m}n_\infty^2\pi H R^2_\mathrm{v}.
\label{eq:EnergyVortexUnperturbed}
\end{equation}

In order to now determine the frequencies of the different oscillation modes, we can either write down the Hamilton-Jacobi equations, or compare \eqref{eq:EnergyVortexUnperturbed} to the Hamiltonian of the harmonic oscillator for $A(t)$ (since we expanded $A(t)$ only up to second order). The frequencies of the oscillation modes are then given by:
\begin{equation}
\begin{aligned}
\omega^2_{l=0}&=\frac{c^2}{2\xi^2}\left[\ln\left(\frac{R_\mathrm{cond}}{R_\mathrm{v}}\right)\right]^{-1},\\
\omega^2_{l\neq 0}&=l\frac{c^2}{2\xi^2}\left[1-\left(\frac{R_\mathrm{v}}{R_\mathrm{cond}}\right)^{2l}\right]^{-1},
\end{aligned}
\label{eq:FrequenciesOscillationModes}
\end{equation}
where $c=\frac{\hbar}{\sqrt{2}m\xi}$ is the speed of sound. Note that in the limit $\xi\rightarrow 0$ (also using \eqref{eq:MinimalRadiusVortexCore}) all frequencies \eqref{eq:FrequenciesOscillationModes} go to infinity, making these oscillation modes energetically inaccessible. For increasing values of $\xi$ all of the frequencies \eqref{eq:FrequenciesOscillationModes} decrease.

\subsection{The axial Kelvin modes (or kelvons)}
\label{subsec:KelvinModes}

The Kelvin modes (or kelvons) have already been extensively studied in BECs. To ensure the vortex is confined  by an infinite potential well, Dirichlet boundary conditions are imposed, resulting in a quantization of the wavenumber for the Kelvin modes:
\begin{equation}
k_n=n\frac{\pi}{H}\qquad\mathrm{with}\qquad n\in\mathbb{N}_0.
\label{eq:kValuesDiricletBoundary}
\end{equation}
The Kelvin modes have a characteristic long-wavelength dispersion relation given by\cite{Thomson,Sonin,Fetter}:
\begin{equation}
\nu_n=\frac{\hbar k_n^2}{2m}\ln\left(\frac{1}{|k_n|\xi}\right)=n^2 c\xi\frac{\pi^2}{H^2\sqrt{2}}\ln\left(\frac{H}{n\pi\xi}\right),
\label{eq:FrequencyKelvinModes}
\end{equation}
where for the last equation the quantization condition \eqref{eq:kValuesDiricletBoundary} was used. The long wavelength approximation for the Kelvin modes breaks down when the wavelength of the Kelvin modes becomes comparable to the healing length. As variations in the condensate wavefunction must occur on length scales larger than the healing length (in order to be collective excitations), this length scale sets an upper bound on the allowed $k_n$ values \eqref{eq:kValuesDiricletBoundary} (and thus $n$ values):
\begin{equation}
k_n\xi<1 \Leftrightarrow \frac{H}{n\pi\xi}>1.
\label{eq:kValuesUpperBound}
\end{equation}
Note from \eqref{eq:FrequencyKelvinModes} that the frequencies (and thus the energies) of the Kelvin modes become smaller as the condensate height $H$ increases.

\subsection{Comparison of the radial oscillation modes and the axial Kelvin modes}
\label{subsec:Comparison}

\begin{figure}[h!tb]
	\centering
	\includegraphics{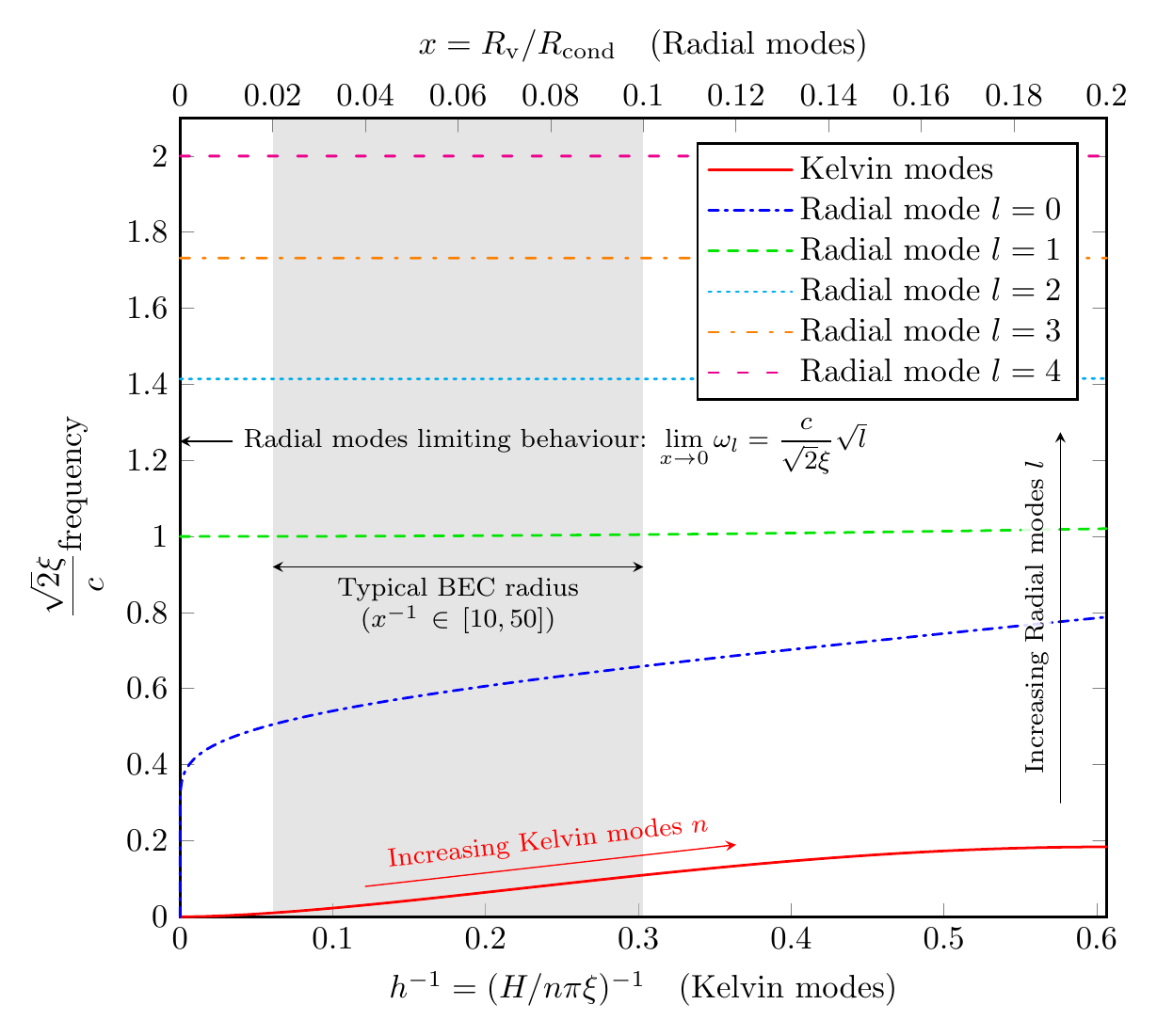}
	\caption{The frequencies of the Kelvin modes \eqref{eq:FrequencyKelvinModes} (solid line) and the first five radial modes \eqref{eq:FrequenciesOscillationModes} (other lines). Both oscillation modes are plotted in terms of the dimensionless variables \eqref{eq:DimensionlessVariables}, the Kelvin modes as a function of the condensate height $h$ (given by the lower axis) and the radial modes as a function of the condensate radius $x$ (given by the upper axis). The gray area represents the range of the typical radius of a Bose-Einstein Condensate (BEC) of atoms ($x^{-1}\in[10,50]$) \cite{Cooper}.}
	\label{fig:FrequencyPlotsComparison}
\end{figure}

The energies of the Kelvin and radial modes can be compared by comparing their frequencies, in other words by solving $\omega_l=\nu_n$. When both frequencies are equal, a resonance between the two different oscillation modes will occur. In order to further simplify the calculations, the dimensionless variables
\begin{equation}
h=\frac{H}{n\pi\xi}\qquad\mathrm{and}\qquad x=\frac{R_\mathrm{v}}{R_\mathrm{cond}}
\label{eq:DimensionlessVariables}
\end{equation}
can be used, where $h$ is the dimensionless condensate height and $x$ is the dimensionless inverse condensate radius. Note that $x\in[0,1]$ and according to \eqref{eq:kValuesUpperBound} $h\in[1,\infty[$. Using the dimensionless variables, the frequencies of the radial \eqref{eq:FrequenciesOscillationModes} and Kelvin modes \eqref{eq:FrequencyKelvinModes} can be plotted, this is done in figure \ref{fig:FrequencyPlotsComparison}.
In figure \ref{fig:FrequencyPlotsComparison} the dimensionless radius $x$ \eqref{eq:DimensionlessVariables} is constrained to the interval $[0,0.2]$. The reason for this constraint is that the model for the vortex core \eqref{eq:SingleVortexUnperturbed} is only valid for a small vortex radius (compared to the condensate radius).

On figure \ref{fig:FrequencyPlotsComparison} the typical vortex to condensate size ratio's for a vortex in a BEC ($x\in[1/50,1/10]$) are indicated by a shaded area. As can be seen, in this range of $x$ values, the radial modes will not show a resonance with the Kelvin modes. Note however that in this $x$ range, the frequencies of the lower radial modes and the Kelvin mode are in the same order of magnitude.

\section{Conclusions}
\label{sec:conclusions}
From our calculations it can be seen that the radial modes for the vortex core will show no resonances with the Kelvin modes within the typical BEC setup ($R_\mathrm{cond}/R_\mathrm{v}\in[10,50]$). This can be seen in figure \ref{fig:FrequencyPlotsComparison}, where the frequencies of the radial and Kelvin modes are plotted in terms of condensate height and radius. 

This means that for low energy decay, even in ultracold gases where the vortex core has a finite size, the Kelvin modes will still have a preferred role. Note that although a strict resonance between the lower radial modes and the Kelvin modes is not possible, both modes have a frequency within the same order of magnitude. This means that including the radial modes may result in a (small) correction when included in calculation.

From figure \ref{fig:FrequencyPlotsComparison} it might be tempting to state that in superfluid helium, where the vortex radius is practically zero, a resonance between the radial and Kelvin modes might be possible. Note however that liquid helium (with\footnote{Where $n$ is the superfluid density and $a_s$ is the s-wave scattering length.} $n|a_s|^3\approx 0.5-2500$) is not a dilute superfluid (where $n|a_s|^3<<1$ should hold). This means that the (mean field) s-wave scattering potential used in \eqref{eq:GrossPitaevskiiEnergy} no longer yields an accurate description of the problem. The mean field  GP description used here is no longer appropriate for superfluid helium, meaning that no conclusions can be drawn for liquid helium from figure \ref{fig:FrequencyPlotsComparison}.

Finally it can be noted that the above study does not take finite temperature effects into account. For a finite temperature it is to be expected that both the Kelvin and radial mode energies are broadened over an energy interval of the size $k_BT$. This thermal blurring will lead to a (slightly) larger region in which both oscillation modes can couple.

\begin{acknowledgement}
The authors acknowledge the weekly fruitful discussions with S.N. Klimin, G. Lombardi, J.P.A. Devreese and W. Van Alphen. This research was supported by the University Research Fund (BOF) of the University of Antwerp, project: 2014BAPDOCPROEX167 and FFB1550168, and by the Flemish Research Foundation (FWO-V1), project nrs: G.0115.12N, G.0119.12N, G.0122.12N and G.0429.15.N.
\end{acknowledgement}

\end{document}